\documentclass[sigconf,nonacm]{acmart}

\acmConference[ICSE 2024]{46th International Conference on Software Engineering}{April 2024}{Lisbon, Portugal}

\AtBeginDocument{%
  \providecommand\BibTeX{{%
    Bib\TeX}}}

\setcopyright{acmcopyright}
\copyrightyear{2023}
\acmYear{2023}
\acmDOI{XXXXXXX.XXXXXXX}

\usepackage{graphicx}
\usepackage{tabularx}
\usepackage{enumitem}

\usepackage{subfig}

\usepackage{listings}
\usepackage{xcolor}
\usepackage{framed}
\usepackage{multirow}
\usepackage{changepage}

\definecolor{codegreen}{rgb}{0,0.6,0}
\definecolor{codegray}{rgb}{0.5,0.5,0.5}
\definecolor{codepurple}{rgb}{0.58,0,0.82}

\lstdefinestyle{mystyle}{
  keywordstyle=\color{magenta},
  numberstyle=\tiny\color{codegray},
  stringstyle=\color{codepurple},
  basicstyle=\ttfamily\footnotesize,
  breakatwhitespace=false,         
  breaklines=true,                 
  captionpos=b,                    
  keepspaces=true,                 
  numbers=left,                    
  numbersep=5pt,                  
  showspaces=false,                
  showstringspaces=false,
  showtabs=false,                  
  tabsize=2
}

\newboolean{showcomments}
\setboolean{showcomments}{true}
\ifthenelse{\boolean{showcomments}}
 { \newcommand{\mynote}[2]{
      \fbox{\bfseries\sffamily\scriptsize#1}      {\small$\blacktriangleright$\textsf{\emph{#2}}$\blacktriangleleft$}}}
        { \newcommand{\mynote}[2]{}}

\definecolor{formalshade}{rgb}{0.85,1,0.85}
\definecolor{darkblue}{rgb}{0.0,0.6,0.30}

\newenvironment{formal}{%
  \MakeFramed{\advance\hsize-\width\FrameRestore}%
  \noindent\hspace{-4.55pt}
  \begin{adjustwidth}{}{}%
}
{%
  \end{adjustwidth}\endMakeFramed%
}

\begin{document}

\title{Automatic Generation of Test Cases based on Bug Reports: \\a Feasibility Study with Large Language Models}

\author{Laura Plein, Wendkûuni C. Ouédraogo, Jacques Klein, Tegawend\'e F. Bissyand\'e}
\email{firstname.lastname@uni.lu}
\affiliation{%
  \institution{University of Luxembourg}
  \country{Luxembourg}
}


\acmArticleType{Research}

\begin{abstract}
Software testing is a core discipline in software engineering where a large array of research results has been produced, notably in the area of automatic test generation. Because existing approaches produce test cases that either can be qualified as simple (e.g., unit tests) or that require precise (and executable) specifications, most testing procedures still rely on test cases written by humans to form development project test suites. 
Such test suites, however, are incomplete: they only cover parts of the project or they are produced after the bug is fixed and therefore can only serve as regression tests. Yet, several research challenges, such as automatic program repair, and practitioner processes, such as continuous integration, build on the assumption that available test suites are sufficient. There is thus a need to break existing barriers in automatic test case generation.
While prior work largely focused on random unit testing inputs, we propose to consider generating test cases that  realistically represent complex user execution scenarios, which reveal buggy behaviour. Such scenarios are informally described in bug reports, which should therefore be considered as natural inputs for specifying bug-triggering test cases. 
In this work, we investigate the feasibility of performing this generation by leveraging large language models (LLMs) and using bug reports as inputs. 
Our experiments consider various settings, including the use of ChatGPT, as an online service for accessing an LLM, as well as CodeGPT, an existing code-related  pre-trained LLM that was fine-tuned for our task. Our study is carried out on the Defects4J dataset. Overall, we experimentally show that bug reports associated to up to 50\% of Defects4J bugs can prompt ChatGPT to generate an executable test case. We show that even new bug reports (i.e., previously-unseen data to mitigate data leakage threat to validity), can indeed be used as input for generating the executable test cases. Finally, we report experimental results which confirm that LLM-generated test cases are immediately useful in software engineering tasks such as fault localization as well as patch validation in automated program repair.

\end{abstract}


\maketitle

\section{Introduction}
Tests suites are a key ingredient in various software automation tasks. Recently, several studies~\cite{liu2019you,smith2015cure,yu2019alleviating} have demonstrated that they are paramount in the adoption of latest innovations in software engineering, such as automated program repair (APR)~\cite{goues2019automated}. APR is indeed nowadays a well-researched field, where various techniques and approaches are proposed to automatically generate bug-fixing patches towards reducing debugging and fixing time. Prominent approaches in the literature systematically require precise specifications of correct/incorrect behaviour, such as test cases, to drive the localization as well as the validation of generated patches. This requirement is further exacerbated as recent studies~\cite{le2018overfitting,le2019reliability,tian2022predicting} have shown that the correctness of generated patches is dependent on the quality of the test suites. Extensive test suites with high coverage are indeed required to assess whether the generated patch is actually fixing the bug without introducing new ones. 

Test suites are unfortunately often too scarce in software development projects~\cite{le2019reliability,xiong2018identifying}. Generally, they are provided for regression testing, while new bugs are discovered by users who then describe them informally in bug reports. In recent literature, a new trend of research in APR has attempted to leverage bug reports in generate-and-validate pipelines for program repair. Approaches such as iFixR~\cite{koyuncu2019ifixr} then target the recommendation of patches instead of systematic application on the buggy code. Yet, when an APR tool generates a patch candidate, if test cases are unavailable,  developers must manually validate the patch, leading to a threat to validity, as recognized in the evaluation of CapGen~\cite{wen2018context}.

Test suites are therefore essential in APR~\cite{yang2017better}. On the one hand, automatic test generation approaches in the literature~\cite{nebut2006automatic,fraser2011evosuite,pacheco2007randoop}, unfortunately, either target unit test cases and thus do not cater to the need for revealing complex bugs that users actually face in the execution of software, or require formally-defined inputs such as the function signatures, or even the test oracle. On the other hand, bug reports are pervasive, but remain under-explored.
There is thus a need to investigate the feasibility of test case generation by leveraging bug reports. Our ultimate objective indeed is to address a challenge in the adoption of program repair by practitioners, towards ensuring that patches can be automatically generated and validated for bugs that are reported by users. By filling the gap between test case generation and bug reports, we expect to establish a game-changing setting towards the adoption of program repair in industry.

Concretely, we observe that, while bug reports can quickly be overwhelming (in terms of high quantity and/or low quality) for developers, they are still recognized to contain a wealth of information. Unfortunately, such information hidden in natural language informality can be difficult to extract, contextualize and leverage for specifying program executions.
Nevertheless, recent advances in Natural Language Processing (NLP) have opened up new possibilities in software engineering. 
In particular, with the advent of large language models (LLMs), a wide range of tasks have seen machine learning achieve, or even exceed, human performance. Machine translation~\cite{lopez2008statistical,stahlberg2020neural} in particular has been a very active field where several case studies have been explored beyond language translation. For example in software engineering, several research directions have investigated the feasibility of leveraging natural language inputs for producing  programming artefacts and vice-versa. Some milestones have been recorded in the literature in code summarization~\cite{allamanis2018survey,hu2018deep}, program repair~\cite{goues2019automated,monperrus2018automatic}, and even program synthesis~\cite{gulwani2017program}. Nevertheless, bug reports have scarcely been explored. Yet, automating bug reproduction via analysis of bug reports holds tremendous value.

\textbf{This paper.} In this work, we propose to {\em study the feasibility of exploiting LLMs towards producing executable test cases based on informal bug reports}. Our experiments build on ChatGPT~\cite{brown2020language} and codeGPT~\cite{DBLP:journals/corr/abs-2102-04664}. The former has recently received much attention and presents the advantage that its model has been trained on a large corpus of natural language text as well as source code of software programs. The latter is a pre-trained model targeting software engineering, and which can be fine-tuned for various tasks.




\textbf{Contributions.} 
The main contributions of this study are :
\begin{itemize}
    \item \emph{Exploration:} We present a pioneering research discussion on the feasibility of automatically generating executable test cases based on user-written bug reports. The main contribution is the assessment of the capabilities of current LLMs for this newly defined task. 
    \item \emph{Findings:} We conduct a comprehensive empirical study based on the Defects4J dataset and explore different experimental settings of ChatGPT and codeGPT. The experimental results yield various findings on the executability, validity and relevance ratios of the generated test cases. We also provide a preliminary analysis of the quality requirements of bug reports as well as of the complexity of the generated test cases to support our initial assumption that bug reports could be relevant inputs for producing realistic test cases. Finally, we experimentally show that the generated test cases will be indeed instrumental in an APR pipeline: they enable accurate fault localization and support the validation of patch correctness.
    \item \emph{Dataset:} Our study produces artefacts that we share with the community to enable future research in this axis. The linked bug reports, the fine-tuned codeGPT model, the time-stamped generated test cases by ChatGPT (v 3.5) as well as the labeled results are made publicly available.
\end{itemize}

In the remainder of this paper we present our experimental setup (Section~\ref{setup}), discussing the benchmarks, dataset, evaluation metrics and present the research questions. Section~\ref{results} presents the results from our empirical study, Section~\ref{discussion} overview the threats to validity and limitations followed by the related work (Section~\ref{rw}) before the conclusion in Section~\ref{conclusion}.
\section{Experimental setup}\label{setup}
Our experimental setup is framed around the need to address bugs that are discovered by users after the software has been shipped. Such new bugs are getting reported on code repositories, such as GitHub, every day, slowing down users and leading to development costs for bug investigations by the project owners. 

\begin{figure}[h!]
  \includegraphics[width=\linewidth]{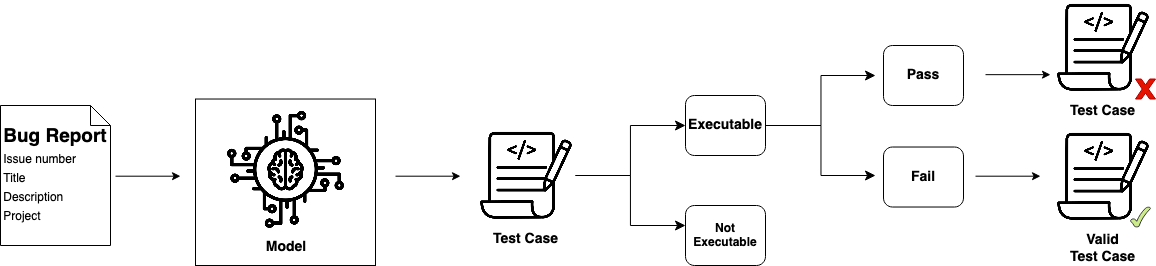}
  \caption{Pipeline for LLM-based generation of test cases from bug reports}
  \label{fig:main}
\end{figure}

To train a model that is able to generate the required test case, our pipeline (see Figure~\ref{fig:main}) includes the following steps: (1) We start by identifying Java projects included and collecting their bug reports; (2) then, we use LLMs (either an online service or a pre-trained model that we fine-tune) for the purpose of generating test cases; (3) once the test cases are generated, they are appended to the existing test suite of the project to assess their \textit{executability} and \textit{validity} as well as their \textit{relevance} for the associated reported bug; (4) finally, once we generated a relevant test case for the given bug, APR tools can now be applied towards generating and validating the bug-fixing patch.

\subsection{Research Questions}
To evaluate the feasibility of translating informal bug reports into executable test cases we would like address the following research questions:

\begin{description}
\item \textbf{RQ1: Can informally-written bug reports be translated into executable test cases with ChatGPT?} To investigate this research question, ChatGPT, a state-of-the-art LLM as-a-service platform, is applied to translate bug reports from Defects4J into test cases.

\item \textbf{RQ2: To what extent can fine-tuned models compare to the baseline results achieved with ChatGPT?} We investigate the capabilities of models fine-tuned for the test case generation task. The generated test cases are compared to those of ChatGPT in terms of executability, validity and relevance (for bug triggering).

\item \textbf{RQ3: Can LLMs actually generate executable test cases for new bugs?} The Defects4J benchmark may have leaked in the trained LLMs. Therefore, we assess the feasibility  of generating test cases for newly reported bugs to validate the generalizability of the translation models.

\item \textbf{RQ4: How does the bug report quality influences the test case generation performance?} The considered bug reports from Defects4J come from different projects with various types of users. While some users write detailed bug reports, others are unprecise and short. With this RQ, we investigate whether the size/content of the bug report has an influence on the capability of the LLM to generate executable/relevant test case.

\item \textbf{RQ5: How relevant are the generated test cases for software engineering tasks?} In this RQ we investigate the impact of our generated test cases on fault localization and patch validation.
\end{description}

\subsection{Benchmark and Dataset}\label{data}
The performance of test case generation with LLMs is assessed based on the Defects4J repository~\cite{defects4J-dissection} which includes real-world faults from various Java software development projects. We collect the bug reports as well as the failing test cases associated to these faults, for every bug. Defects4J additionally provides the buggy and the fixed project version. This benchmark is well suited for this study since it has been widely used to evaluate state-of-the-art APR tools as well as in the software testing community. 

To answer RQ1, all available bug reports were used to generate test cases. One must mention that some bugs referred to the same bug report, in that case it was only considered once to avoid bias in the results because of duplicates. For RQs that require fine-tuning the LLM for the task of test case generations, pairs of bug reports and failing test cases are required. Eventually, the constructed dataset contains 972 pairs of bug reports and failing test cases as detailed in Table~\ref{table:defects4j}.

\begin{table}[h!]
\begin{center}
\resizebox{\columnwidth}{!}{%
\begin{tabular}{ |c|c|c|c|c| c| c|} \cline{2-7}
\multicolumn{1}{c|}{}  & \multicolumn{6}{c|}{\bf Projects}   \\  \cline{2-7}
 \multicolumn{1}{c|}{}  & Chart  & Closure & Lang & Math & Mockito & Time  \\ 
\hline
\bf  \# of pairs of & \multirow{2}{*}{14} & \multirow{2}{*}{370} & \multirow{2}{*}{161} & \multirow{2}{*}{187} &  \multirow{ 2}{*}{174} & \multirow{2}{*}{66} \\ 
\bf Bug Report \& Test Case &  &  &  &  &  &  \\ 
 \hline
\end{tabular}%
         }
\centering
\caption{Dataset composition}
 \label{table:defects4j}
\end{center}
\end{table}

\subsection{Baselines}
For this study we consider different baselines from the literature and from latest open releases. We focused on models that were already used for the task of natural language to code translation:  we used \textbf{ChatGPT} directly as baseline to generate test cases while we opted to fine-tune \textbf{CodeGPT}~\cite{DBLP:journals/corr/abs-2102-04664} specifically for our task. 
On the one hand, the ChatGPT API facilitates its integration into the APR pipeline. On the other hand, CodeGPT has been successfully applied to various NLP tasks, including in code-to-code, code-to-NL and NL-to-code tasks. CodeGPT further provides a pre-trained model adapted to Java code.

\subsection{Prompt Design for ChatGPT}\label{prompt}
To query ChatGPT for generating a test case based on a bug report, we used the following simple prompt design: we concatenate two pieces of information: the instruction and the bug report. The instruction is unique for all queries to ChatGPT and is as follows: ``{\textit{write a Java test case for the following bug report: ...}}''. For the bug report, our feasibility study considers that no pre-processing should be applied on the bug report. However, the submitted information should not include follow-up comments or attachments.

\subsection{Fine-Tuning CodeGPT}\label{finetuning}
CodeGPT~\cite{DBLP:journals/corr/abs-2102-04664} is a generative pretrained model that was fine-tuned on Java source code to be applied for NL-to-code tasks. Various model versions of CodeGPT are available in the literature repositories. To fine-tune CodeGPT to translate the bug reports into test cases (NL-to-code) we used the dataset that we prepared from the Defects4J faults as mentioned in Section~\ref{data}. After the fine-tuning, CodeGPT was used to translate the unseen bug reports into Java test cases.


\subsection{Metrics}
\subsubsection{\textbf{Test case evaluation}}
To evaluate the quality of the generated test cases, we rely on the following metrics.\\
\begin{itemize}
\item\textbf{Executability} is a binary metric which describes whether the test case is directly executable on the corresponding project version without any manual changes.
\item {\bf Validity: } an executable test case may or may not fail on the target buggy program. We follow the convention of patch validation in program repair and consider the generated test case to be valid only when it, indeed, fails on the buggy program. Otherwise it is considered as invalid. 
\item\textbf{Relevance} is another binary metric which describes if a {\em valid} test case can not only reproduce the bug but also validate the patched version. To verify the relevance of a test case for a given bug we have the two following criteria. First, it need to be valid. Additionally, the test case must pass on the fixed project version. Only if both criteria are met, we consider the generated test case as relevant for the given bug report. 
\end{itemize}
\subsubsection{\textbf{Bug Report Quality evaluation}} To evaluate the impact of the bug report quality, approximated by the size and the presence of code examples, on the performance of test case generation (in terms of executability and relevance), we rely on the \textbf{Mann-Whitney U} also called Mann-Whitney-Wilcoxon (MWW)~\cite{mcknight2010mann} test. In order to apply the MWW metric and calculate the p-value, the size of each bug report was represented by the amount of characters of the bug report.  The p-value (between 0 and 1) of the MWW test measures the statistical significance of the observed difference between two groups (in our case bug reports leading to executable/relevant or to non executable/irrelevant test cases). This means that a small p-value suggests the evidence against the null hypothesis, meaning that there is a significant difference between two groups, while a large p-value indicates that there is no significant evidence of difference.
\label{sec:metrics}

\subsubsection{\textbf{Patch Validation}}
To evaluate the generated patches we will follow the evaluation of the TBar~\cite{liu2019tbar} paper. Thus, we will define a patch as plausible if the project version with the generated patch passes all the tests from the test suite.

\section{Experimental Results}\label{results}
This section presents, for each research question, the experimental objective, the experimental design and implementation details, as well as the yielded results.

\begin{figure*}[!t]
  \begin{minipage}[b]{0.3\linewidth} 
    \centering
    \includegraphics[width=\linewidth]{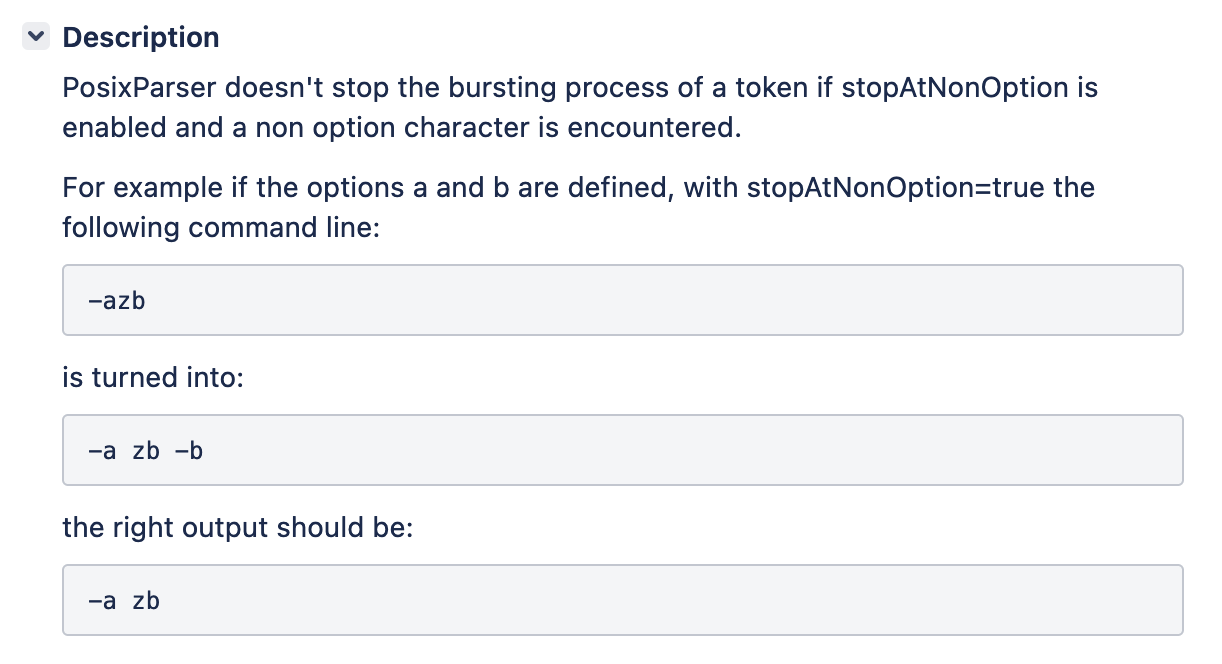}
  \caption{Example bug report from Defects4J dataset - Cli 17}
    \label{fig:gencli}
  \end{minipage}
  \hfill 
  \begin{minipage}[b]{0.3\linewidth} 
\begin{lstlisting}[language=Java, frame=single]
public void testPosixParserStopAtNonOption() throws ParseException {
    String[] args = {"-azb"};
    Options options = new Options();
    options.addOption("a", false, "Option A");
    options.addOption("b", false, "Option B");
    CommandLineParser parser = new PosixParser();
    CommandLine cmd = parser.parse(options, args, true);

    assertTrue(cmd.hasOption("a"));
    assertTrue(cmd.getOptionValue("a").isEmpty());
    assertTrue(cmd.getArgs()[0].equals("zb"));
    assertFalse(cmd.hasOption("b"));
}
\end{lstlisting}
    \captionof{lstlisting}{ChatGPT-generated test case for Cli 17}
    \label{lst:generated}
  \end{minipage}
  \hfill
  \begin{minipage}[b]{0.3\linewidth}
  \begin{lstlisting}[language=Java,frame=single]
public void testStopBursting() throws Exception
    {
        String[] args = new String[] { "-azc" };

        CommandLine cl = parser.parse(options, args, true);
        assertTrue( "Confirm -a is set", cl.hasOption("a"));
        assertFalse( "Confirm -c is not set", cl.hasOption("c"));

        assertTrue( "Confirm  1 extra arg: " + cl.getArgList().size(), cl.getArgList().size() == 1);
        assertTrue(cl.getArgList().contains("zc"));
    }
\end{lstlisting}
\captionof{lstlisting}{Original test case associated to Cli 17}
    \label{lst:original}
  \end{minipage}
\end{figure*}

\subsection{[RQ1]: LLM baseline performance on test case generation with ChatGPT}
\noindent\textbf{[Experiment Goal]:} The goal is to assess whether  test cases produced by human developers could have been generated with an off-the-shelf LLM using reports written in natural language for informally describing the bugs. To answer this first research question, we will explore two sub-RQs considering ChatGPT (version 3.5) as a baseline model. We perform two distinct experiments: a first experiment considers the generation of a single test case to demonstrate the feasibility of our approach; a second experiment performs multiple generations of test cases, based on recent findings in the literature~\cite{tian2023chatgpt}.

\noindent\textbf{[Experiment Design] (RQ1.1 - single generation):} 
For our experiments we rely on the ChatGPT API (version 3.5) and used the default parameters. In order to use ChatGPT to generate test cases, the prompt given to the API consists of two parts as introduced in Section~\ref{prompt}. The ChatGPT API was then used to generate the test cases for all the bugs in the study subjects.
In practice, before running the generated test cases, the ChatGPT outputs are parsed to clean them from natural language texts (e.g., descriptive details) that would lead to compilation failures. Afterwards, the test cases are systematically included in the test suite, which is fully executed by the Defects4J test pipeline. Execution results are then logged, allowing us to compute the metrics on executability, validity and relevance.

\noindent\textbf{[Experiment Results]  (RQ1.1 - single generation):} 
Figure~\ref{fig:gencli} provides an illustrative example of a bug report (from the CLI project) and the associated test case (Listing~\ref{lst:original}) (ground truth in Defects4J) and the generated test case (Listing~\ref{lst:generated}) from ChatGPT. As we can see in this example, ChatGPT is able to generate an executable test case from a bug report that can be used to reproduce it. This can enable various software automation tasks, such as spectrum based fault localization, patch validation in program repair, and more generally automated software testing. As we can see in Figure~\ref{fig:gencli} and in Listing~\ref{lst:generated}, ChatGPT was able to extract important semantic parts from the bug report such as relevant inputs and function names.

On the Defects4J dataset, we compute the proportion of bug reports for which ChatGPT is able to successfully generate test cases. We evaluate the amount of generated test cases that are directly executable without the need of further developer changes. Then we evaluate the percentage of valid and relevant test cases.
Note that in Table~\ref{fig:genPerc} the amount of bug reports is the amount of distinct bug reports, which does not necessarily match the amount of bugs identified in the Defects4J dataset for a given project. Indeed, we found that in most projects there were several bugs that were associated to the same bug report.

\begin{table*}[!t]
\begin{center}
\resizebox{0.9\textwidth}{!}{
\begin{tabularx}{1\textwidth}{  
  | >{\centering\arraybackslash}X 
  || >{\centering\arraybackslash}X
  || >{\centering\arraybackslash}X
  | >{\centering\arraybackslash}X 
  | >{\centering\arraybackslash}X 
  | >{\centering\arraybackslash}X 
  | >{\centering\arraybackslash}X 
  | |>{\centering\arraybackslash}X 
  | >{\centering\arraybackslash}X 
  | >{\centering\arraybackslash}X 
  | >{\centering\arraybackslash}X 
  | >{\centering\arraybackslash}X | } 
 \cline{3-12}
 \multicolumn{1}{c}{} & \multicolumn{1}{c}{} & \multicolumn{5}{|c||}{\bf single generation attempt}  & \multicolumn{5}{c|}{ \bf multiple (5) generation attempts} \\ 
 \cline{1-12}
 Project & \# of bug reports & overall exec.  & overall valid. &  valid. over exec. & overall relev. & relev. over exec.&  overall exec. &  overall valid.  &  valid over exec. & overall relev. & relev. over exec.\\ 
\hline
 Chart & 6 & 0\% & 0\%& 0\%& 0\% & 0\% & 33\% &17\% &50\%&17\% &50\%\\ 
 Cli & 30 & 37\% & 23\%& 64\% &10\% & 27\% & 53\% &37\%& 69\% &17\% &31\%\\ 
 Closure & 127 &  6\%& 3\% & 50\% & 3\% &  50\% & 46\% & 28\%& 59\%&3\% &7\%\\ 
 Lang & 60 & 25\% &10\% &40\%& 17\% &67\% & 60\% &43\%&72\% & 27\% &44\%\\ 
 Math & 100 & 13\% &3\%&23\%& 3\% &23\% & 43\% &15\% &35\% &6\% & 14\%\\ 
 Time & 19 & 32\% &32\%&100\%& 0\% &0\% & 84\% &17\%& 81\% &0\% &0\%\\ 
 \hline
  \hline
 Total & 342 & 15\% & 7\%&  47\% & 6\% & 38\% & 50\% & 30\% & 59\%& 9\% & 19\% \\
 \hline
\end{tabularx}
}
 \caption{Performance of ChatGPT on the task of test case generation based on bug report. \\{\normalfont Legend: exec. $\rightarrow$ executability; valid. $\rightarrow$ validity. relev. $\rightarrow$ relevance} }
 \label{fig:genPerc}
\end{center}
\end{table*}



As depicted in Table \ref{fig:genPerc}, with a single generation attempt, we reached the highest percentage (37\%) of executable test cases for the Cli project. The validity of the generated test cases varies greatly from one project to another, which could imply an influence of the quality of the bug reports. As mentioned in Section~\ref{data}, every project had a different source and format of user-written bug reports. Therefore, their quality may differ significantly across projects. We explore the influence of quality in RQ4.

On average we reached an executability of 15\% when generating a single  test case per bug report. This result confirms that ChatGPT is indeed able to generate test cases using  user-written bug reports as prompts. Only 6\% of the total amount of generated test cases in this experiment were, however, relevant. While this proportion appears limited, it confirms the feasibility of using LLM for the task of test generation. Moreover, we noted that 38\% of the executable test cases are actually relevant, which is promising, since executability is a binary metric that is easy to automatically infer. While, overall, only 7\% of the generated test cases were valid, it is noteworthy that 47\% of the executable ones are valid. It is further important to mention that due to the randomness\footnote{we used the default value of the \textit{temperature} parameter, which controls the randomness of the generated text.} of ChatGPT generation, only one generation might not be representative of the full potential of the model.

Experimental challenges: an issue we encountered with some Math project versions is that the test suite is compiling but their execution is never ending. Therefore, there were 10 Math bug reports for which we could not determine the relevance of the generated test cases.

Additionally, the low results on the relevance might come from the fact that ChatGPT was directly leveraged to generate test cases without any fine-tuning on the task at hand, and therefore, it potentially had no specific knowledge about the projects' context (in terms of what the test target is) but only some general knowledge on Java syntax.

\noindent\textbf{[Experiment Design] (RQ1.2 - multiple generations):}
RQ1.1 proved the feasibility of generating executable and relevant test cases with ChatGPT, using the bug report as prompt. Due to the randomness of ChatGPT, the generated test cases quality might differ significantly from one generation to the next. Therefore, an additional experiment was done to investigate if at least one out of five generated test cases could be executable and relevant. 
In this experiment, for every prompt, we query the ChatGPT API five (5) times and assess the different generated test cases. 

\noindent\textbf{[Experiment Results]  (RQ1.2 - multiple generations):} 
In this section we discuss and compare the results received while generating one versus 5 test cases, following the experimental procedure of a recent work~\cite{tian2023chatgpt}. According to the data in Table~\ref{fig:genPerc}, on average, executable test cases were obtained for 50\% of the bug reports across all projects, while we obtained at least one valid test cases for 30\% of bug reports. Furthermore, among the executable test cases, we observed that 59\% were valid. Further manual investigations highlighted that \textit{executability}, \textit{validity} and \textit{relevance} can, in a large number of cases, be fixed with minor modifications (e.g. adding relevant package imports or changing duplicate function names). Overall, the obtained results  confirm that ChatGPT is a promising tool in the generation of test cases based on bug reports. 


The fact that we gain +35\% points of executable, +23\% points of valid and +3\% points of relevant test cases as described in Table \ref{tab:genComp}, shows that once the initial \textit{executability} challenge is passed, the tests generated by ChatGPT are actually valid, highlighting its capability of capturing semantics of bug reports, towards translating those bug reports into bug-triggering test cases. These feasibility study results strongly motivate further research in the area. 

\begin{figure}[!h]
    \centering
    \includegraphics[width=0.5\linewidth]{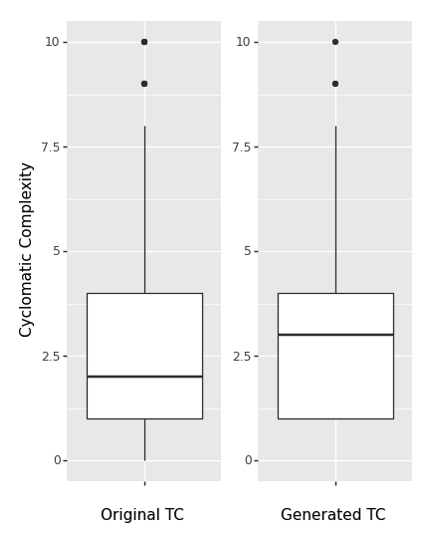}
    \caption{Cyclomatic complexity of ground-truth bug-triggering test cases (Original TC) vs LLM-generated valid test cases (Generated TC)}
    \label{fig:cc_all}
\end{figure}

We provide in Figure~\ref{fig:cc_all} the distributions of the cyclomatic complexity~\cite{gill1991cyclomatic} values of the generated valid test cases compared to that of the original (ground truth) bug-triggering test cases provided by developers. The difference of median values suggest that test cases generated from bug reports tend to be more complex. Our postulate is that such test cases attempt to fit with the detail inputs for reproducing the reported bug, while developers write test cases that focus to the actual key bug-triggering input. 
 
\begin{formal}
\textbf{Answer to RQ1:} \textit{Our experimental results show that ChatGPT can be prompted with bug reports to generate executable test cases for 50\% of the input samples. Beyond \textit{executability},  about 30\% of the bugs could be reproduced with valid test cases, and about 9\% of all generated test cases were actually relevant. Nonetheless, we note that over half (59\%) of the executable test cases were valid test cases. These results, which are based on an off-the-shelf LLM as-a-service, show promises for automated test case generation, leveraging complex information from user-reported bugs.}
\end{formal}

\subsection{[RQ2]: Performance of CodeGPT, a code-specific LLM fine-tuned on the task of test case generation}\label{rq2}

\noindent\textbf{[Experiment Goal]:} To evaluate to what extent fine-tuned models performance can compare to the baseline results obtained with ChatGPT, we resort to fine-tuning CodeGPT.

\noindent\textbf{[Experiment Design] (RQ2.1 - single generation):}\label{codegpt} 
For our experiments, we consider the codeGPT-small-java-adaptedGPT2 model, which we fine-tune on the Defects4J Java test cases following the default parameters for fine-tuning. To train a model that can generalize in the best possible way, we randomly shuffled the Defects4J dataset  since the bug reports of every project came in different format, size and quality. The dataset was eventually split into 70\% training, 15\% testing and 15\% validation.

\noindent\textbf{[Experiment Results] (RQ2.1 - single generation):} Table~\ref{tab:genComp} summarizes the experimental results for this RQ. In a single generation attempt, CodeGPT outperforms ChatGPT: the fine-tuned CodeGPT generated 24\% of executable test cases. The results suggest that fine-tuning a generative pretrained model on the specific task of translating natural language bug reports into  test cases results in a better performance than simply leveraging ChatGPT in a single generation attempt. Since the fine-tuned CodeGPT model is based on the GPT version 2.0, We expect the results to be substantially better once the GPT 4.0 models will be available for fine-tuning.

Among the executable test cases, 20\% were also relevant for the given reported bug. This performance clearly motivates the fine-tuning of GPT models, given that they will gain project-specific knowledge, which is essential for generating a large proportion of not only executable but also relevant test cases. 

\noindent\textbf{[Experiment Design] (RQ2.2 - multiple generations):} For this experiment we used the previously fine-tuned CodeGPT model (Section~\ref{codegpt}) and applied it to make five times generations on the testing dataset while changing the random seed from one experience to the next to get different results. The generated test cases were again evaluated on executability, validity and relevance.

\noindent\textbf{[Experiment Results] (RQ2.2 - multiple generations):} As shown in Table~\ref{tab:genComp}, performing multiple generations with a fine-tuned model only slightly increase the number of executable test cases. Fine-tuned models are by nature less random in their generations compared to leveraging pre-trained models as ChatGPT. Additionally, the training dataset was quite small which might have led to overfitting. This explains the only slight increase (+10\% points) in executable test cases when generating multiple test cases per bug report with CodeGPT compared to an increase of +35\% points of executable test cases when performing multiple generations with ChatGPT.\\

\begin{table}[!t]
\begin{center}
\resizebox{\columnwidth}{!}{%
\begin{tabular}{ |c||c|c||c|c|c|} 
 \cline{2-5}
 \multicolumn{1}{c||} {}   &  \multicolumn{2}{c||}{single generation}  &  \multicolumn{2}{c|}{5 generations} \\ 
\hline
     & ChatGPT  & CodeGPT  & ChatGPT & CodeGPT\\ 
\hline
  Executability of all test cases  & 15\%  & 24\%  & 50\% & 34\%   \\ 
  \hline
    Validity  of all test cases & 7\%  & 15\%  & 30\% & 17\%   \\ 
  \hline
Validity of executable test cases   & 47\%  & 60\% & 59\% & 51\%    \\ 
  \hline
  Relevance of all test cases &  6\% &  5\% & 9\% & 6\% \\ 
 \hline
  Relevance of executable test cases  &  38\% & 20\%  & 19\% & 17\%  \\ 
 \hline
\end{tabular}}
 \caption{Performance of fine-tuned CodeGPT vs ChatGPT on the task of bug report driven test case generation}
 \label{tab:genComp}
\end{center}
\end{table}

\begin{formal}
{\bf Answer to RQ2:}
\textit{Fine-tuning CodeGPT yielded an LLM that generates executable test cases for 24\% of the bug reports with a single generation attempt. This rate is substantially larger than the one achieved by the ChatGPT single generation baseline (15\%). However, when performing several generation attempts, ChatGPT achieves a significantly higher rate of success. These results suggest that a fine-tuned LLM could be beneficial in an automated pipeline where a single shot is adequate, whereas ChatGPT would be more useful in recommendation scenarios. The results further suggest that more powerful models should be investigated in future work.}
\end{formal}

\subsection{[RQ3]: Performance generalization on new bugs}
\noindent\textbf{[Experiment Goal]:} 
We leveraged, for our experiments, ChatGPT and CodeGPT, two LLMs that are known to have been trained on public data (up to October 2021). Since we rely on the Defects4J dataset, a widely used benchmark in testing and program repair, it is likely that some samples from this dataset have been included in the models' training data. This constitutes a threat to the validity of our results as it reflects a potential data leakage problem. To address this concern, and to reliably confirm the feasibility of the idea, we evaluate the proposed generation pipeline on newly reported bugs (after October 2021), where we can guarantee that no test cases that reproduce the bugs were part of the models' original training data.

\noindent\textbf{[Experiment Design]:} Our experiment is focused on ChatGPT (multiple generation), and evaluates executability automatically. Since there is not yet a fixed version for all new bugs, a manual investigation of the relevance of the generated test case is necessary. We collect all available Defects4J bug reports that were created after the 1st of October 2021 to be sure that they were not included in the training data of ChatGPT 3.5. There were only new bugs reported for the Cli, Lang and Math projects. We therefore considered additional projects, which are still maintained. 

\noindent\textbf{[Experiment Results]:} 
Table~\ref{fig:newExec} indicates the number of new bugs that were assessed as well as the performance of ChatGPT in terms of executability. Within 3 projects from the previous study and 2 additional projects, we collected 38 new bug reports: for 55\% of those bug reports, the generated test cases were executable. This performance is on par with the performance achieve with old bug reports of Defects4J (cf. RQ2), suggesting that the previously-obtained performance  is likely unbiased. Furthermore, we note that a large proportion (50\% for Lang and 100\% for JacksonDatabind) of executable generated test cases are valid. 
We manually analyse relevance and confirm the feasibility of generating relevant test cases. In the absence of an oracle (fixed version to assess relevance of the test case), we make available our dataset for the community to build on.

\begin{table}[!t]
\begin{center}
\resizebox{\linewidth}{!}{
\begin{tabular}{ |c|c|c|c|c| } 
 \hline
  Project & \# new bugs & \# executable  & \% executable & \% valid over exec. \\ 
\hline
  Cli & 3 & 0 &  0\% & - \\ 
 Lang & 12 & 8 &  67\% & 50\%\\ 
 Math & 5 & 2 &  40\% & - \\ 
 JacksonDatabind & 5 & 4 &  80\% & 100\%\\ 
 Jsoup & 13& 7& 54\% & - \\
 \hline
 \textbf{Total} & 38 & 21& 55\% & 47\%\\ 
 \hline
\end{tabular}
}
 \caption{Executability for newly reported bugs}
 \label{fig:newExec}
\end{center}
\end{table}





\begin{figure}[!t]
\begin{lstlisting}[frame=single]
Posted also on StackOverflow: https://stackoverflow.com/questions/74917912/url-encoding-in-jsoup-not-working-properly

When I changed version from 1.11.3 to 1.15.3 I started getting MalformedUrlException when fetching URLs with characters that need encoding, like: https://im-creation-assets.s3-us-west-2.amazonaws.com/CelebrityCars[DE]/20221208JuliaRobertsCarJvo/juliayoung-1___native_1200-627.jpg)
Because of the '[' and ']' in the URL...

The exception comes from org.jsoup.helper.CookieUtil#asUri - and was added somewhere between those versions.

I see that your code tries to encode in org.jsoup.helper.HttpConnection#encodeUrl - but this encoding does not work on this URL (and many more).
I can do the encoding myself before calling org.jsoup.Jsoup#connect - the problem is that if there is a redirect to such a URL, I get back to this error.

We do encoding, by the way, using springframework, and it works pretty well, something like:

import org.springframework.web.util.UriComponentsBuilder;
final String encodedUrl = UriComponentsBuilder
.fromUriString(url)
.build()
.encode()
.toUri()
.toString();

(I know I can avoid redirects - do them myself - and encode every URL... but maybe it is possible to fix this encoding issue for everyone...).

Thanks!
\end{lstlisting}
\captionof{lstlisting}{New bug report from JSoup (Dec 26, 2022)}
    \label{lst:new_jsoup_br}
\end{figure}

Listing~\ref{lst:new_jsoup_br} provides an example of a newly reported  bug in Jsoup that we use to prompt ChatGPT for generating the bug triggering test case (i.e., for reproducing the associated bug). Note that this bug report is still unresolved and that no associated test case has been provided by the user nor by other users (when it was again submitted on stackoverflow). 
With ChatGPT, we generated an executable test case, as shown in   Listing~\ref{lst:generated_jsoup_new}. A manual assessment further confirmed that inputs extracted by ChatGPT to build the test case are relevant and help to reproduce the reported bug.

\begin{figure}[!t]
\begin{lstlisting}[language=Java, frame=single]
import java.io.IOException;
import java.net.HttpURLConnection;
import java.net.URL;
import java.net.URLEncoder;
import java.nio.charset.StandardCharsets;
import org.junit.Test;
import static org.junit.Assert.assertEquals;

@Test
public void testUrlEncoding() throws IOException {
    String originalUrl = "https://im-creation-assets.s3-us-west-2.amazonaws.com/CelebrityCars[DE]/20221208JuliaRobertsCarJvo/juliayoung-1___native_1200-627.jpg";

    // Encode the URL manually before fetching
    String encodedUrl = URLEncoder.encode(originalUrl, StandardCharsets.UTF_8.toString());

    URL url = new URL(encodedUrl);
    HttpURLConnection connection = (HttpURLConnection) url.openConnection();

    // Set the request method if necessary (GET is the default)
    connection.setRequestMethod("GET");

    // Get the response status code
    int statusCode = connection.getResponseCode();

    // Check if the response status code is successful (2xx) to confirm that the URL was fetched without exceptions
    assertEquals(true, statusCode >= 200 && statusCode < 300);
}
\end{lstlisting}
    \captionof{lstlisting}{Generated Jsoup Test Case}
    \label{lst:generated_jsoup_new}
\end{figure}

\begin{formal}
{\bf Answer to RQ3:} \textit{ChatGPT, an LLM-as-a-service, was proven capable of generating executable test cases for newly reported bugs. Overall, in over 55\% cases, the new bug reports could effectively serve as prompts to generate an executable test case. Through manual analysis, we confirmed that the generated test case reflects the described behaviour.}
\end{formal}

\subsection{[RQ4]: Impact of Bug Report Quality}
\noindent\textbf{[Experiment Goal]:} In this experiment, we approximate bug report quality with its size as well as the presence of code artefacts. This approximation is based on prior findings in the literature: very short bug reports are often of low quality (i.e., they do not contain sufficient information for reproducibility), while bug reports that include code excerpts are often written by developers who provide enough details for reproduction. We therefore investigate the impact of the bug report size and content on the performance of the test case generation. For the content we will distinguish between bug reports containing code snippets and bug reports that include only natural language text. 

\noindent\textbf{[Experiment Design] (RQ4.1 - bug report size):} We consider the categories of bug reports based on whether they led to the generation of executable, valid and relevant test cases with ChatGPT or not. Then we computed the size (number of characters of the bug report) of all bug reports per category.

\begin{figure}[!t!]
\centering
  \includegraphics[width=\linewidth]{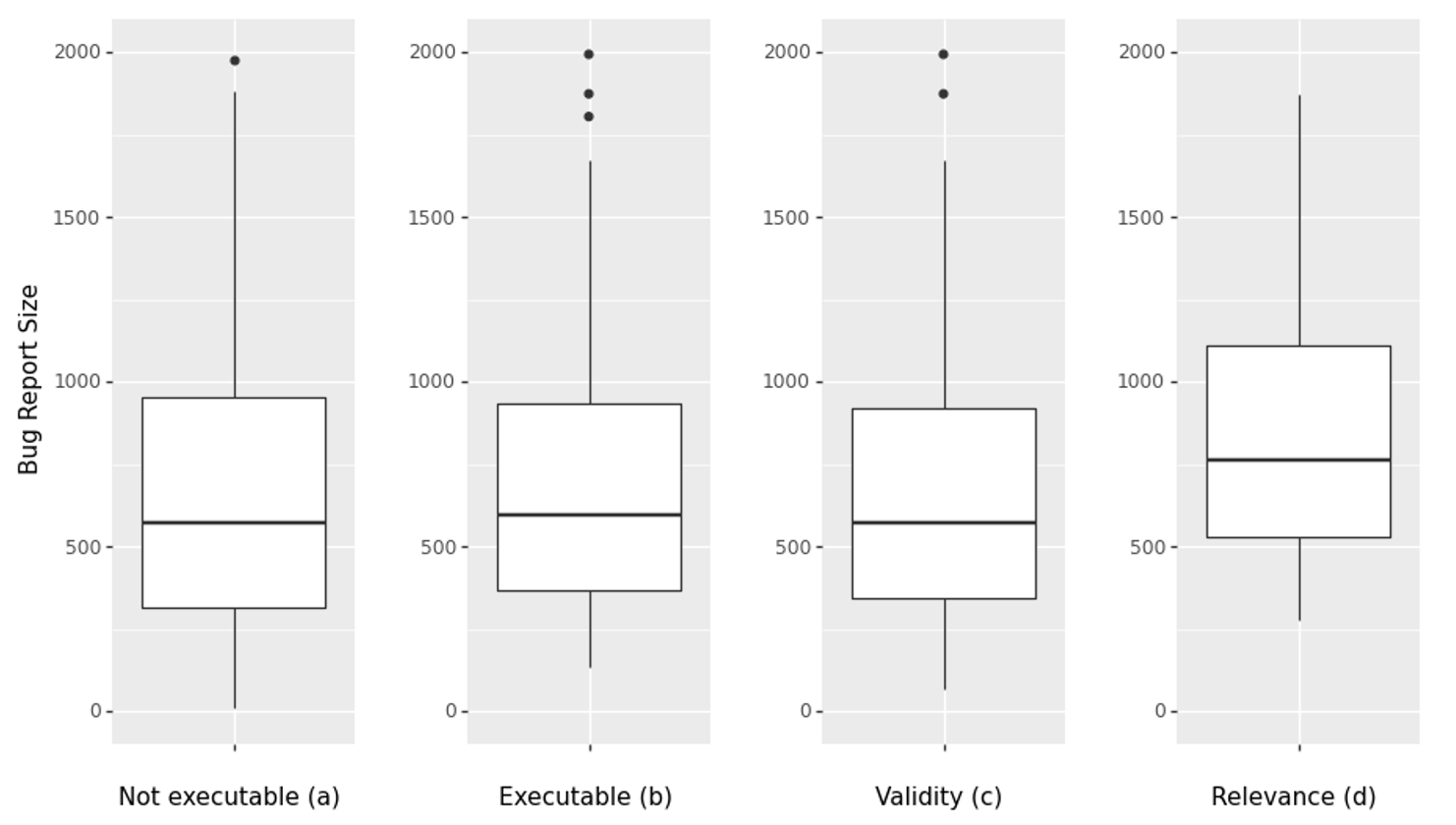}
  \caption{Bug report size impact on generated test cases}
  \label{fig:br_sizes}
\end{figure}

\noindent\textbf{[Experiment Results] (RQ4.1 - bug report size):} The median values of bug report size distributions per category, as presented in Figure~\ref{fig:br_sizes}, suggest that bug report size has little influence over the executability and validity of the generated test case. In contrast, relevant test cases are associated with bug reports with larger sizes. The statistical significance of the differences between sets was evaluated based on the Mann Whitney U score. With  a p-value that is largely > 0.5 (0.99), we could conclude that the bug report size has no impact on the performance of test generation in terms of executability.

This observation is potentially related to the fact that ChatGPT has been extensively trained on publicly available Java source code, enabling it to generate test cases with minimal syntax errors. Consequently, the executability and validity of the generated test cases remains largely independent of the bug report size. 


The influence of the bug report size on the performance in terms of relevance appears to be a more important aspect to assess since syntactic correctness of the generated test cases is not enough. The MWW test indeed confirms that the bug report size has a slight impact : p-value <0.5 (0.2).

\noindent\textbf{[Experiment Design (RQ4)] (RQ4.2 - bug report content):} We also consider the presence of code snippets within the bug reports and assess its impact on the performance of the test case generation. Code presence is considered confirmed if at least one parsable code statement is included. Thus, the simple mention of a function name in a sentence will not count as code in this experiment.



\begin{figure}[h!]
\centering
  \includegraphics[width=\linewidth]{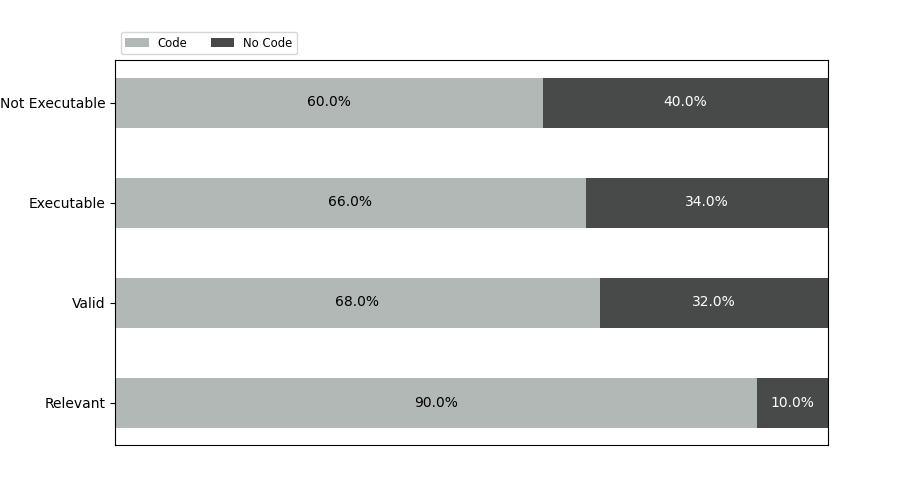}
  \caption{Impact of Code presence in bug report on the generated test cases}
  \label{fig:rep_content}
\end{figure}

\noindent\textbf{[Experiment Results] (RQ4.2 - bug report content):} 
Our initial observations of the data from RQ1 hint to the fact that the quality of the bug reports matters when generating test cases using ChatGPT with bug reports as prompts.
Figure \ref{fig:rep_content} represents the status of the bug report in terms of code presence as well as the evaluation of the generated test cases for datasets involving the Cli, Closure, Lang, Math and Time project samples

On average, 63\% of all the bug reports in our dataset contained code. Therefore, as shown in Figure \ref{fig:rep_content} the number of bug reports which include code is always higher than the number of bug reports which do not. Nevertheless, the data clearly reveal that the better the test case is (i.e., executable, then valid, and finally relevant), the higher the probability that the associated input bug report contains code. If we compare the percentages of bug reports leading to the generation of non-executable vs those leading to executable test cases, we already observe an increase of 6\% points in terms of code presence. Going from executable to valid test cases, there is only a slight increase in code presence (2\% points). However, for the relevance, 90\% of the bug reports leading to relevant test cases contained code. This clearly highlights the impact of code snippets within the bug report on the generation task by LLMs. Our assumption is that such code snippets provide a much-needed context for the target context of the test case.



\begin{formal}
{\bf Answer to RQ4:} \textit{The experimental results suggest that the bug report size has little impact on the executability and validity of the generated test cases. However, it has a slight influence on the probability for the test case to be relevant.
With respect to the content (and the presence of code in the bug report),  the experiments revealed that 66\% of the bug reports leading to the generation of executable test cases contained code, while 90\% of relevant test cases are associated to bug reports which contain code. These observations confirm that the code present in the natural language bug report serves as an important guidance to the LLM for generating test cases that are executable and relevant.}
\end{formal}


\subsection{[RQ5]: Usages in software engineering}
\begin{figure}[h!]
\centering
  \includegraphics[width=\linewidth]{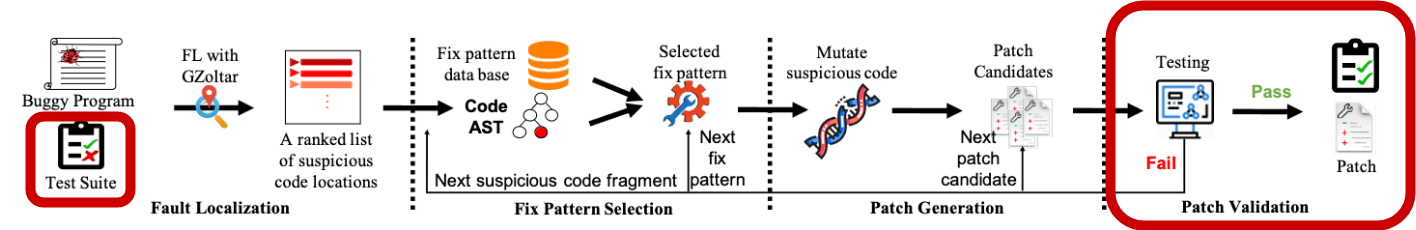}
  \caption{Typical generate-and-validate APR pipeline (e.g., TBar~\cite{liu2019tbar})}
  \label{fig:tbar}
\end{figure}

\noindent\textbf{[Experiment Goal]:} In this experiment we investigate how relevant the generated test cases are for software engineering tasks such as fault localization and patch validation in automated program repair. The following experiments will demonstrate the feasibility of using  test cases generated from user-written bug reports in order to run a full generate-and-validate repair pipeline in production. As shown in Figure~\ref{fig:tbar}, the classical pipeline requires test cases to successfully localize bugs (RQ5.1 - Fault Localization) and validate the generated repair patches (RQ5.2 - Patch Validation).

\noindent\textbf{[Experiment Design] (RQ5.1 - Fault Localization):} In a first experiment we  evaluate that fault localization performance can be ensured with the generated test cases. Following the classical steps in the APR literature, we rely on spectrum-based fault localization (SBFL)~\cite{liu2019you}, and evaluate the results obtained by running GZoltar\footnote{https://gzoltar.com/} (version 1.7.4) with Ochiai, for every project in our study where we were able to generate a valid test case. Running GZoltar for every buggy project version provides a list of all suspicious source code lines ranked by the Ochiai suspiciousness score. We consider a bug as localized if at least one of the actual buggy source code line is in the top ranking of this list. For this study we will distinguish between top 1 and top 5 ranking. 
In a first step we run GZoltar on the original test suite (i.e., with the ground truth test cases) which will provide our reference performance. In a second step we will run GZoltar on the test suite where we removed the original bug-triggering test case and replaced it with our valid generated test case.

\noindent\textbf{[Experiment Results] (RQ5.1 - Fault Localization):} Table~\ref{table:sbfl} summarizes the spectrum-based fault localization results. On the original test suite, our SBFL implementation was able to perfectly localize the buggy source code line (top 1 ranked) for 22 bugs in our dataset. When considering the top 5 outputs of the SBFL, 42 bugs were correctly localized in the reference setting. 

\begin{table}[t!]
\begin{center}
\resizebox{\linewidth}{!}{
\begin{tabularx}{0.5\textwidth}{  
  | >{\centering\arraybackslash}X 
  | >{\centering\arraybackslash}X
  | |>{\centering\arraybackslash}X
  | >{\centering\arraybackslash}X 
  | |>{\centering\arraybackslash}X 
  | >{\centering\arraybackslash}X | } 
 \cline{3-6}
 \multicolumn{1}{c }{} &\multicolumn{1}{c|| }{}  & \multicolumn{2}{c||}{w/ ground truth} & \multicolumn{2}{c|}{w/ generated} \\ 
\multicolumn{1}{c }{}  &\multicolumn{1}{c|| }{}  & \multicolumn{2}{c||}{test cases} & \multicolumn{2}{c|}{test cases} \\ 
 \hline
  Project& \# Bugs & GZ 1 Line &  GZ 5 Line & GZ 1 Line &  GZ 5 Line  \\ 
\hline
 Closure & 35 &4 &7 &\textbf{5}  &\textbf{10}\\ 
 Lang & 25 & 10 & 18& \textbf{11} &16\\ 
 Math & 15 & 6&  13 & \textbf{7} &\textbf{14}\\ 
 Time & 13 &2 &  4 & 1 &4\\ 
 \hline 
 Total & 88 &22 &42 &\textbf{24}  &\textbf{44}\\ 
 \hline 
\end{tabularx}
}
 \caption{Number of bugs localized with GZoltar/Ochiai}
 \label{table:sbfl}
\end{center}
\end{table}

Interestingly, however, when running the SBFL on a test suite including the generated test cases (from bug reports) instead of the ground truth test cases, the localization performance is slightly improved: the reference results are improved for 3 of the study projects, in terms of localization both @top-1 and @top-5. Overall, the bug-report based generated test cases where able to help precisely localize 4 bugs that could not be localized even with the ground truth test cases: 2 additional bugs at top 1 and 2 additional bugs at top 5. Given that the generated test cases are more complex than  the ground truth ones (cf. RQ1), it is possible that it facilitates a better discrimination of relevant code lines with spectrum-based fault localization. 


\noindent\textbf{[Experiment Design] (RQ5.2 - Patch Validation):} In a second experiment we evaluate to what extent the generated test cases can help validate automatically-generated patches. We consider patches generated by the TBar~\cite{liu2019tbar} template-based APR baseline tool. TBar was selected as it merges fix patterns templates from various works, and it has achieved a high performance on the Defects4J benchmark. This tool systematically attempts to fix each bug by iteratively applying its fix patterns and validating the patched program using the project test suite. Despite this repair technique being simple, it remains among the best performing in terms of the number of bugs fixed~\cite{liu2020efficiency}. 
We then investigate how many TBar-generated patches that are labeled as correct (resp. plausible) are validated by the generated test cases (i.e., they fail on the unpatched program and pass on the patched ones). 
Table~\ref{table:intersection} summarizes the intersection of projects used for this study. \\

\begin{table}[h!]
\begin{center}
\resizebox{\linewidth}{!}{
\begin{tabularx}{0.5\textwidth}{  
  | >{\centering\arraybackslash}X 
  | >{\centering\arraybackslash}X
  | >{\centering\arraybackslash}X 
  | >{\centering\arraybackslash}X | } 
 \hline
Project  & Correct (Plausible) TBar Patches & Valid Test Cases & Intersection   \\ 
\hline
 Closure & 17 (9) & 35 & 6 (0) \\ 
 Lang & 13 (5) & 26 & 6 (4) \\ 
 Math & 22 (12) & 15 & 1 (2) \\ 
 Time & 3 (3) & 13 & 0 (2) \\ 
 \hline 
\end{tabularx}
}
 \caption{Intersection of bugs with correct patches and valid test cases}
 \label{table:intersection}
\end{center}
\end{table}


\noindent\textbf{[Experiment Results] (RQ5.2 - Patch Validation)):}
Table~\ref{table:intersection} summarizes the results. It shows for example that among the 9 patches of Closure that were known to be plausible (i.e., passing the project test suite, but being actually incorrect), none of them were found valid by the generated test cases. Overall, only 8 out of the 29 plausible patches were validated by the generated test cases. This means that the generated test cases could have been more useful than the existing developer-written test suites to discard 21 plausible patches.

Similarly, however, we note that several patches that were manually labeled in the TBar dataset as correct did not get validated by the generated test cases. Since we have validated the generated test case by ensuring that it does not fail on the patched version of the program, it is possible that the manual labeling (based on judgement of semantic equivalence of patches) led to some mistakes, or that the actual developer test cases were  incomplete in the scoping of the reported bug, or that the generated test case is rather overfitting the bug report. In any case, our experimental results trigger relevant research questions for future work in bug report driven test case generation. Note that TBar did not generate any patches for the Cli project, and for the Chart project there were no intersecting faults for which we had correctly generated patches and valid generated test cases. 
Overall, the results show a promising research direction in the feasibility of using test cases generated from user bug reports in order to validate generated patches, enabling the adoption of generate and validate APR tools in real-world software development cycles.\\



\begin{formal}
{\bf Answer to RQ5: }\textit{The experiments with spectrum-based fault localization have shown that running test cases generated by LLMs using bug reports as inputs can lead to even better localization performance than with ground truth test cases. Additionally, we have shown that the generated test cases can indeed be leveraged to validate patch correctness.}

\end{formal}
\section{Discussion}\label{discussion}
Due to the randomness of \textbf{ChatGPT}, it is essential to verify that ChatGPT is correctly replying to the prompt before starting all the experiments. Meaning that one prompt might lead to the generation of test cases in Java source code one day but lead to the generation of explanations of test cases in natural language another day.

Overall, our empirical study validates the feasibility of using LLMs for automatic test case generation based solely on informal bug reports. However, our proposed pipeline bears some limitations while our empirical results carry some threats to validity. We enumerate those in this section and discuss some research directions for future work to increase the amount of executable and valid test cases among the LLM-generated ones.

\subsection{Threats to validity}

Our \textbf{Dataset}, inferred from Defects4J data and used to fine-tune CodeGPT is relatively small, which can easily lead to overfitting. It would be beneficial to investigate additional projects to collect more bug reports and failing test case pairs. Additionally, the dataset only contains Java test cases, which was easy for a first study on the feasibility of generating test cases from bug reports. However, this focus affects the generalisation of the results. In future work, it would be relevant to investigate the feasibility of generating test cases for programs written in other programming languages.

In RQ5.2 the \textbf{patch validation} is carried out for patches generated by a single APR tool, namely TBar. Since the intersection of valid generated test cases and correctly or at least plausibly fixed patches by a single APR tool is quite small it would be most relevant, in future work, to assess the performance of the LLM-generated test cases on patches generated by different APR tools.

\subsection{Limitations \& Future Work}
In this study the \textbf{Executability} only reflects if a generated test case is directly executable or not. This doesn’t reflect the amount of effort for a human to make it executable. After manually reviewing the generated test cases, we saw that most can be made executable through the modification of one or two lines of code. The most common issues are usually: missing imports, duplicate function names, or the use of a deprecated function. Those limitations could systematically be fixed in future work (e.g. with prompt engineering) significantly increasing the amount of executable test cases.

In our experiments \textbf{bug reports} where collected and directly used as prompt to demonstrate the feasibility and applicability of our idea to address real software faults reported by the users. Nevertheless, pre-processing the textual data might be beneficial to keep the LLM focused on the main context of the bug report, therefore improving the validity and relevance of the generated test case.

In this study we used the \textbf{default parameters} while querying ChatGPT as well as for fine-tuning CodeGPT. Future work should consider tuning some parameters such as the temperature in order to potentially reach a higher amount of executable and relevant test cases.

Many powerful \textbf{Pre-trained Language Models} where recently released, which are not yet available for fine-tuning, those should also be considered in future work. In future work, further experiments with other models such as codeBERT, PLBART, RoBERTa or the LLaMA~\cite{touvron2023llama} model could also be done to determine which model is the most suited for the task. The experiment in Section~\ref{rq2} with CodeGPT gave us 24\% of executable test cases. A manual investigation gave insights on some of the failing reasons, one simply being that the test case was not generated until the end and thus, was syntactically incorrect. This issue never happened with ChatGPT. Therefore, we strongly believe that fine-tuning more powerful models, will result in a much higher executability and relevance rate since it will combine the projects knowledge through the fine-tuning process and it will have a solid basic knowledge not to do any syntactic errors.





\section{Related Work}\label{rw}
Writing extensive test suites, i.e., which have a high code coverage is extremely costly in terms of time and developer expertise. Several techniques~\cite{anand2013orchestrated,taneja2008diffgen,thummalapenta2009mseqgen,fraser2013evosuite} have therefore been proposed in the literature to help developers automatically generate unit tests for new software units. To address the lack of formal specifications for generating test cases, Fischbach et al.~\cite{fischbach2023automatic} proposed an NLP-based approach for defining informal requirements to the software. This example study shows that some first experiments have been performed on investigating informal sources for test case generation even though their work still requires a strict natural language structure containing conditionals for their requirements. In contrast, in our study, we are aiming at using the unprocessed human written bug report as direct input for test case generation. Recently, some work on test case generation using large language models (LLM) was done by~\cite{schafer2023adaptive}, but their TestPilot still requires the functions signature and implementation as prompt. However, we consider that no formal input from the source code is required. Instead, only the informal bug report is required, towards implementing an approach that would be generalizable and software independent. With the recent success of ChatGPT many researchers are investigating it's potential in software engineering tasks. A recent study~\cite{yuan2023no} has proven ChatGPTs capabilities of generating unit tests but for their study they considered the classes source code as prompt. Feng et al.~\cite{feng2023prompting} have investigated ChatGPTs ability to help developers reproduce the bug while extracting important steps to reproduce the bug from the bug report. However they did not perform any test case generation. The existing test suite of a project has a significant impact on the quality of the generated patches by an APR tool as investigated by Liu et al.~\cite{liu2020efficiency}. This highlights the importance of being able to generate additional test cases to guide the software repair tools. APR tools do not always generate correct and optimal patches for a given bug~\cite{le2018overfitting,tian2022predicting}, therefore, augmenting the test suite is essential to help reduce the over-fitting issue. The use of ChatGPT to directly enhance APR techniques ~\cite{xia2023automated,xia2023conversational,sobania2023analysis} highlights even more the potential of this new LLM.
\section{Conclusion}\label{conclusion}
In this work, we confirmed the feasibility of {\em 
automating test case generation based on bug reports using large language models.}
Basic experimental results using ChatGPT have indeed suggested that this LLM-as-a-service, despite not beeing fine-tuned for the specific task of test case generation, is also able to successfully use bug reports (written in natural language) as prompts to generate executable and relevant bug-triggering test cases. 

Subsequent experiments, showed that while ChatGPT achieves high executability with multiple generations, the fine-tuned LLMs (ex: CodeGPT) achieve a higher number of executable test cases than ChatGPT with a single generation attempt. These findings suggest that, while promising, the research agenda on generating test cases with LLMs still has a performance gap to close. Nevertheless, additional experiments on newly reported bugs have suggested that the achieved baseline performance is genuine: it is not biased by training data leakage. 

Finally, investigations on the relevance of the LLM-generated test cases in software engineering have demonstrated their potential for fault localization as well as patch validation, two key steps in automated program repair. Overall, we showed that despite the simplicity of the general translation idea, LLMs offer substantial concrete opportunities in an open problem of software engineering research which is {\em how to automatically construct relevant test suites}. These test suites are key ingredients not only in traditional testing research,  but also towards the practical adoption of generate-and-validate techniques in automated program repair. 

\vspace{0.5cm}
\noindent
{\bf Open Science.} We make available all our artefacts (code, datasets and results) in the following repository:
\begin{center}
    \url{https://anonymous.4open.science/r/BR-to-Test-Case-7237}
\end{center}

\bibliographystyle{ACM-Reference-Format}
\bibliography{references.bib}

\end{document}